\documentclass[twocolumn]{aastex701}
\usepackage[T1]{fontenc}
\usepackage{graphicx}
\usepackage[]{natbib}
\usepackage{amsmath, multirow,hhline}

\usepackage[colorinlistoftodos]{todonotes}

\usepackage[percent]{overpic}

\newcommand{\hst}{\textit{HST}}
\newcommand{\jwst}{\textit{JWST}}

\newcommand{\photutils}{\texttt{Photutils}}

\newcommand{\nifty}{\texttt{NIFTY}}
\newcommand{\bagpipes}{\texttt{Bagpipes}}
\newcommand{\galfit}{\texttt{Galfit}}

\newcommand{\astrodrizzle}{\texttt{Astrodrizzle}}

\newcommand{\Mjup}{M$_{\mathrm{Jup}}$}

\usepackage{hyperref}
\begin{document}
\title{Two Exciting High-redshift Galaxy Candidates Turn Out to Be Two Exciting Brown Dwarfs}
\shorttitle{Two Dropouts Turn Out to Be Two Brown Dwarfs}
\shortauthors{Brada\v{c} et al.}

\author[0000-0001-5984-0395]{Maru\v{s}a Brada{\v c}}
\email[show]{marusa.bradac@fmf.uni-lj.si}  
\affiliation{University of Ljubljana, Faculty of Mathematics and Physics, Jadranska ulica 19, SI-1000 Ljubljana, Slovenia}
\affiliation{Department of Physics and Astronomy, University of California Davis, 1 Shields Avenue, Davis, CA 95616, USA}

\author[0000-0002-4201-7367]{Chris Willott}
\affiliation{NRC Herzberg, 5071 West Saanich Rd, Victoria, BC V9E 2E7, Canada}
\email{placeholder}  

\author[0000-0003-3983-5438]{Yoshihisa Asada}
\affiliation{Dunlap Institute for Astronomy and Astrophysics, 50 St. George Street, Toronto, Ontario M5S 3H4, Canada}
\email{placeholder}

\author[0000-0003-0475-9375]{Lo\"ic Albert}
\affiliation{Institut Trottier de recherche sur les exoplan\`etes and D\'epartement de Physique, Universit\'e de Montr\'eal, 1375 Avenue Th\'er\`ese-Lavoie-Roux, Montr\'eal, QC H2V 0B3, Canada}
\email{placeholder}

\author[0009-0009-4388-898X]{Gregor Rihtar\v{s}i\v{c}}
 \affiliation{University of Ljubljana, Faculty of Mathematics and Physics, Jadranska ulica 19, SI-1000 Ljubljana, Slovenia}
\email{placeholder}

\author[0000-0001-9414-6382]{Anishya Harshan}
\affiliation{University of Ljubljana, Faculty of Mathematics and Physics, Jadranska ulica 19, SI-1000 Ljubljana, Slovenia}
\email{placeholder}

\author[0009-0000-2101-1938]{Jon Jude\v{z}}
\affiliation{University of Ljubljana, Faculty of Mathematics and Physics, Jadranska ulica 19, SI-1000 Ljubljana, Slovenia}
\email{placeholder}  

\author[0000-0003-3243-9969]{Nicholas S. Martis}
\affiliation{University of Ljubljana, Faculty of Mathematics and Physics, Jadranska ulica 19, SI-1000 Ljubljana, Slovenia}
\email{placeholder}

\author[0000-0002-9400-7312]{Andrea Ferrara}
\affiliation{Scuola Normale Superiore, Piazza dei Cavalieri 7, I-56126 Pisa, Italy}
\email{placeholder}  

\author[0000-0003-4565-8239]{Kevin Hainline}
\affiliation{University of Arizona Department of Astronomy and Steward Observatory: Tucson, AZ, US}
\email{placeholder}

\author[0000-0002-5258-8761]{Abdurro'uf} \affiliation{Department of Astronomy, Indiana University, 727 East Third Street, Bloomington, IN 47405, USA}
\email{placeholder}

\author[0000-0003-2718-8640]{Joseph F. V. Allingham}
\affiliation{Department of Physics, Ben-Gurion University of the Negev, P.O. Box 653, Be'er-Sheva 84105, Israel}
\email{allingha@post.bgu.ac.il}

\author[0000-0003-0212-2979]{Volker Bromm}
\affiliation{Department of Astronomy, The University of Texas at Austin, Austin, TX 78712, USA}
\affiliation{Cosmic Frontier Center, The University of Texas at Austin, Austin, TX 78712, USA}
\affiliation{Weinberg Institute for Theoretical Physics, University of Texas at Austin, Austin, TX 78712, USA}
\email{placeholder}

\author[0000-0002-0302-2577]{John Chisholm}
\affiliation{Department of Astronomy, The University of Texas at Austin, Austin, TX 78712, USA}
\affiliation{Cosmic Frontier Center, The University of Texas at Austin, Austin, TX 78712, USA} 
\email{placeholder}

\author[0000-0001-7410-7669]{Dan Coe}
\affiliation{Space Telescope Science Institute (STScI), 3700 San Martin Drive, Baltimore, MD 21218, USA}
\affiliation{Center for Astrophysical Sciences, Department of Physics and Astronomy, The Johns Hopkins University, 3400 N Charles St. Baltimore, MD 21218, USA}
\affiliation{Association of Universities for Research in Astronomy (AURA), Inc.~for the European Space Agency (ESA)}
\email{placeholder}

\author[0000-0001-8325-1742]{Guillaume Desprez} 
\affiliation{Kapteyn Astronomical Institute, University of Groningen, P.O. Box 800, 9700AV Groningen, The Netherlands}
\email{placeholder}

\author[0000-0001-9065-3926]{Jose M. Diego}
\affiliation{Instituto e F\'isica de Cantabria,(CSIC-UC), Avda. Los Castros s/n. 39005, Santander, Spain}
\email{placeholder}

\author[0000-0002-9382-9832]{Andreas L. Faisst}
\email{afaisst@caltech.edu}
\affiliation{IPAC, California Institute of Technology, 1200 E. California Blvd. Pasadena, CA 91125, USA}

\author[0000-0001-7201-5066]{Seiji Fujimoto}
\affiliation{David A. Dunlap Department of Astronomy and Astrophysics, University of Toronto, 50 St. George Street, Toronto, Ontario, M5S 3H4, Canada}
\affiliation{Dunlap Institute for Astronomy and Astrophysics, 50 St. George Street, Toronto, Ontario, M5S 3H4, Canada}
\email{placeholder}

\author[orcid=0000-0001-6278-032X]{Lukas J. Furtak}a
\affiliation{Cosmic Frontier Center, The University of Texas at Austin, Austin, TX 78712, USA} 
\affiliation{Department of Astronomy, The University of Texas at Austin, Austin, TX 78712, USA}
\email{\url{furtak@utexas.edu}}
\email{placeholder}

\author[0000-0003-4512-8705]{Tiger Yu-Yang Hsiao}
\affiliation{Department of Astronomy, University of Texas, Austin, TX 78712, USA}
\affiliation{Cosmic Frontier Center, The University of Texas at Austin, Austin, TX 78712, USA}
\email{placeholder}

\author[0000-0001-9840-4959]{Kohei Inayoshi} \affiliation{Kavli Institute for Astronomy and Astrophysics, Peking University, Beijing 100871, China}
\email{placeholder}

\author[0000-0002-6610-2048]{Anton M. Koekemoer} 
\affiliation{Space Telescope Science Institute (STScI), 3700 San Martin Drive, Baltimore, MD 21218, USA} 
\email{koekemoer@stsci.edu}

\author[0000-0002-5588-9156]{Vasily Kokorev}
\affiliation{Department of Astronomy, The University of Texas at Austin, 2515 Speedway, Stop C1400, Austin, TX 78712, USA} 
\affiliation{Cosmic Frontier Center, The University of Texas at Austin, Austin, TX 78712, USA}
\email{vasily.kokorev.astro@gmail.com}

\author[0000-0002-1428-7036]{Brian C. Lemaux}
 \affiliation{Gemini Observatory, NSF NOIRLab, 670 N. A'ohoku Place, Hilo, Hawai'i, 96720, USA}
 \affiliation{Department of Physics and Astronomy, University of California Davis, 1 Shields Avenue, Davis, CA 95616, USA}
 \email{placeholder}
\author[0000-0003-2540-7424]{Paulo A. A. Lopes}
\affiliation{Observatório do Valongo, Universidade Federal do Rio de Janeiro, Ladeira do Pedro Antônio 43, Rio de Janeiro RJ 20080-090, Brazil}
\email{placeholder}

\author[0000-0003-1581-7825]{Ray A. Lucas}
\affiliation{Space Telescope Science Institute (STScI), 3700 San Martin Drive, Baltimore, MD 21218, USA} 
\email{placeholder}

\author[0000-0001-9002-3502]{Danilo Marchesini}
\email{missing@missing.edu}
\affiliation{Department of Physics and Astronomy, Tufts University, 574 Boston Avenue, Suite 304, Medford, MA 02155, USA}

\author[0000-0002-5694-6124]{Vladan Markov}
\email{missing@missing.edu}
 \affiliation{University of Ljubljana, Faculty of Mathematics and Physics, Jadranska ulica 19, SI-1000 Ljubljana, Slovenia}

\author{Gaël Noirot}
 \affiliation{Space Telescope Science Institute, 3700 San Martin Drive, Baltimore, MD 21218, USA}
\email{placeholder}

\author[0000-0002-9651-5716]{Richard Pan}
\affiliation{Department of Physics and Astronomy, Tufts University, 574 Boston Avenue, Suite 304, Medford, MA 02155, USA}
\email{placeholder}

\author[0000-0002-3984-4337]{Scott W. Randall}
\affiliation{Center for Astrophysics, Harvard \& Smithsonian, 60 Garden Street, Cambridge, MA 02138, USA}
\email{placeholder}

\author[0000-0001-5492-1049]{Johan Richard}
\affiliation{Universit\'e Claude Bernard Lyon 1, CRAL UMR5574, ENS de Lyon, CNRS, Villeurbanne, F-69622, France}
\email{placeholder}

\author[0000-0002-6265-2675]{Luke Robbins}
\affiliation{Department of Physics and Astronomy, Tufts University, 574 Boston Avenue, Suite 304, Medford, MA 02155, USA}
\email{placeholder}

\author[0000-0001-8830-2166]{Ghassan T. E. Sarrouh}
\affiliation{Department of Physics and Astronomy, York University, 4700 Keele St., Toronto, Ontario, M3J 1P3, Canada}
\email{gsarrouh@yorku.ca}

\author[0000-0002-7712-7857]{Marcin Sawicki} 
 \affiliation{Department of Astronomy and Physics and Institute for Computational Astrophysics, Saint Mary's University, 923 Robie Street, Halifax, B3H 3C3, Nova Scotia}
\email{placeholder}
\author[0000-0002-6987-7834]{Tim Schrabback}
 \affiliation{Universit\"{a}t Innsbruck, Institut f\"{u}r Astro- und Teilchenphysik, Technikerstr.
 25/8, 6020 Innsbruck, Austria}
 \email{placeholder}
\author[0000-0002-9909-3491]{Roberta Tripodi}
\email{placeholder}
 \affiliation{INAF - Osservatorio Astronomico di Roma, Via Frascati 33, I-00078 Monte Porzio Catone, Italy}
 \affiliation{University of Ljubljana, Faculty of Mathematics and Physics, Jadranska ulica 19, SI-1000 Ljubljana, Slovenia}
 \affiliation{IFPU - Institute for Fundamental Physics of the Universe, via Beirut 2, I-34151 Trieste, Italy}

\author[0000-0002-5057-135X]{Eros Vanzella}
\affiliation{INAF -- OAS, Osservatorio di Astrofisica e Scienza dello Spazio di Bologna, via Gobetti 93/3, I-40129 Bologna, Italy}
\email{placeholder}

\author[0000-0001-8156-6281]{Rogier A. Windhorst}
\affiliation{School of Earth and Space Exploration, Arizona State University,
Tempe, AZ 85287-6004, USA}
\email{Rogier.Windhorst@asu.edu}

% \author[0009-0001-0778-9038]{Giordano Felicioni}
% \affiliation{University of Ljubljana, Faculty of Mathematics and Physics, Jadranska ulica 19, SI-1000 Ljubljana, Slovenia}

% \author[0000-0003-2416-1557]{Douglas Clowe}
% \affiliation{Department of Physics, Ohio University, 1 Ohio University, Athens, OH 45701, USA}

% \author[0000-0002-0933-8601]{Anthony H. Gonzalez}
% \affiliation{Department of Astronomy, University of Florida, Bryant Space Science Center, Gainesville, FL 32611, USA}

% \author[0000-0003-2206-4243]{Christine Jones}
% \affiliation{Center for Astrophysics, Harvard \& Smithsonian, 60 Garden St, Cambridge, MA 02138, USA}

% \author[0000-0003-0144-4052]{Maxim Markevitch}
% \affiliation{NASA/Goddard Space Flight Center, Greenbelt, MD 20771, USA}

 %\author[0000-0002-8530-9765]{Lamiya Mowla}
% \affiliation{Whitin Observatory, Department of Physics and Astronomy, Wellesley College, 106 Central Street, Wellesley, MA 02481, USA}

% \author[0000-0002-8040-6785]{Annika H. G. Peter}
% \affiliation{Department of Physics, Department of Astronomy, and CCAPP, The Ohio State University}
% \author[0000-0002-0086-0524]{Andrew Robertson}
% \affiliation{Carnegie Observatories, 813 Santa Barbara Street, Pasadena, CA 91101, USA}

\def\pmo{$(49 \pm 8)\,\mbox{mas/yr}$}
\def\pmt{$(24 \pm 3)\,\mbox{mas/yr}$}
\def\pmno{$49 \pm 8$}
\def\pmnt{$24 \pm 3$}
\def\pmod{$(430\pm 70)\,\mbox{pc}$}
\def\pmtd{$(860\pm 110)\,\mbox{pc}$}
\def\dir{$20\deg$}

\def\tonerange{272\mbox{--}351\mbox{K}}
\def\ttworange{445\mbox{--}525\mbox{K}}

\def\toneS{324^{+27}_{-26}\mbox{K}}
\def\ttwoS{471^{+24}_{-26}\mbox{K}} 
\def\teffoS{$T_{\rm eff} = \toneS$}
\def\tefftS{$T_{\rm eff} = \ttwoS$}
\def\dteffoS{$240^{+40}_{-40}\,\mbox{pc}$}
\def\dtefftS{$580^{+60}_{-50}\,\mbox{pc}$}

\def\toneA{296^{+12}_{-24}\mbox{K}}
\def\ttwoA{510^{+15}_{-21}\mbox{K}} 
\def\teffoA{$T_{\rm eff} = \toneA$}
\def\tefftA{$T_{\rm eff} = \ttwoA$}
\def\dteffoA{$160^{+20}_{-10}\,\mbox{pc}$}
\def\dtefftA{$580^{+50}_{-70}\,\mbox{pc}$}

\def\bdmodelA{ATMO2020}
\def\bdmodelS{Sonora Elf Owl}
% Abstract of the paper
\begin{abstract}

From the onset of observations of JWST we have discovered unexpectedly luminous galaxies at redshifts $z>10$ and as high as $z=14$. With their discovery, the question immediately followed as to where their progenitors are, since such progenitors should be within reach of existing surveys. However, the discovery of several bright candidates at $z>15$ may indicate further discrepancies between pre-JWST model predictions and current observations. Progenitors of the bright $z\sim 14$ galaxies should be visible at redshifts as high as $z\sim 20-30$, showing in the data as F277W and F356W dropouts. We identify two such candidates in the Bullet Cluster JWST data; however, subsequent NIRSpec follow-up data show spectra that can be well fit with Y dwarf templates with temperatures {\tonerange} and {\ttworange} (using {\tt \bdmodelA} and {\tt \bdmodelS} models) and distances of $\sim 150\mbox{--}650\mbox{pc}$.  The first is one of the lowest-temperature brown dwarfs known, and the lowest-temperature brown dwarf detected spectroscopically outside the solar neighborhood.  With additional NIRCam imaging taken $\sim 1$ year later, we also detect their proper motions of {\pmo} and {\pmt}, further indicating that at least some F277W and F356W dropouts are sub-stellar cold Milky Way objects such as brown dwarfs. 

\end{abstract}


\section{Introduction}
\label{sec:intro}

Recent deep surveys have revealed an unexpectedly large number of bright galaxy candidates at $z>10$, suggesting that significant star formation and stellar mass assembly were already underway within the first few hundred million years after the Big Bang (e.g., \citealp{adamo25} for a review).  The discovery of $z\sim 14$ sources like JADES-GS-z14-0 \citep{carniani24} and MoM-z14 \citep{naidu26} naturally raised a question of where and how can we detect progenitors of these relatively luminous galaxies. 

Several groups have reported photometric candidates at $z>15$ \citep[e.g.,][]{castellano25, perezgonzalez25, asada26, hainline2026zgt8, gandolfi26}. These galaxies are selected by the combination of the dropout technique and/or photometric redshift. At the highest redshifts ($z\sim 20-30$), the Lyman-$\alpha$ break is shifted to $>3\,\mu$m making selection with NIRCam more difficult as galaxies are detected in only a few bands in most extragalactic surveys. With only a few bands, low-redshift solutions invoking a combination of a Balmer break, dust and strong emission lines can mimic high-redshift galaxy spectral energy distributions (SEDs, e.g., \citealp{naidu22}). Furthermore, their small sizes often mean that they are unresolved in NIRCam's long-wavelength channels. 

A recently promising candidate for a galaxy at  $z \sim 32$ is Capotauro \citep{gandolfi25}. This F356W dropout, discovered in CEERS data, is interesting, as it is also detected in NIRSpec follow-up. Its potential variability (detected with $\sim 800$ day baseline, by comparing NIRCam and NIRSpec data) and non-detection of proper motion could even suggest that the object might harbor a pair-instability supernova (PISN) at $z\sim 15$ originating from a massive, metal-free star (\citealt{ferrara26}; see also \citealt{Jeon_PISN2026}). At present, however,  the data cannot unambiguously determine the nature of this object and distinguish if it is indeed located at high redshift or if it might be an extreme low temperature and large distance ($T_{\rm eff}<300\,\mbox{K}$, $130\,\mbox{pc} \lesssim d \lesssim 2\,\mbox{kpc}$) galactic brown dwarf (BD).

Brown dwarfs are important for the characterization of the low-mass end of the stellar initial mass function and represent a link between planet and star formation. Deep extragalactic {\hst} observations have initiated the search for faint BDs (e.g., \citealp{ryan11}), and {\jwst} has now opened a new parameter space in these studies. E.g., T ($500\lesssim T_{\rm eff} \lesssim 1200~\mbox{K}$) and Y-dwarfs ($T_{\rm eff} \lesssim 500\mbox{K}$) have now been found in several extragalactic deep fields with {\jwst} \citep[e.g.,][]{hainline25,burgasser24, morrissey26, li26} to large distances ($\gtrsim 1\mbox{kpc}$). The number densities differ for different surveys, but are roughly $\sim 0.1\,\mbox{arcmin}^{-2}$ for T and $\sim 0.02\,\mbox{arcmin}^{-2}$ for Y dwarfs \citep{hainline25}. For a typical NIRCam pointing, we thus expect roughly 1 T and 0.2 Y dwarfs in the field, though the numbers may vary significantly with galactic latitude and imaging depth.  

In this paper, we present two dropouts from the Bullet Cluster data (a F277W and F356W dropout, corresponding to $z\sim 20-30$ if they are Lyman-break galaxies) that share similar characteristics to Capotauro (and other similar dropouts). The advantage of the data presented here is that together with  NIRCam photometry we obtained NIRSpec follow-up as well as additional NIRCam imaging $\sim 1 \,\mbox{year}$ later, offering a perfect opportunity to study the potential proper motion and variability of these sources. This allows one to characterize them as galactic vs. extragalactic objects. 

This paper is structured as follows. In Section~\ref{sec:data} we
present the data used in this paper and in 
Section~\ref{sec:dataanalysis} we describe the analysis of the photometric and spectroscopic data. In
Section~\ref{sec:results} we present the main science results.  We discuss and summarise our results in
Sections~\ref{sec:discussion} and \ref{sec:conclusions}.
Throughout the paper, we assume a
$\Lambda$CDM cosmology  with $\Omega_{\rm m}=0.3$ and Hubble constant
$H_0=70{\rm\ kms^{-1}\:\mbox{Mpc}^{-1}}$ for ease of comparison with previous work. All uncertainties are given as $1\sigma$ unless otherwise stated.

\begin{figure*}
\begin{overpic}[width=1.0\textwidth]{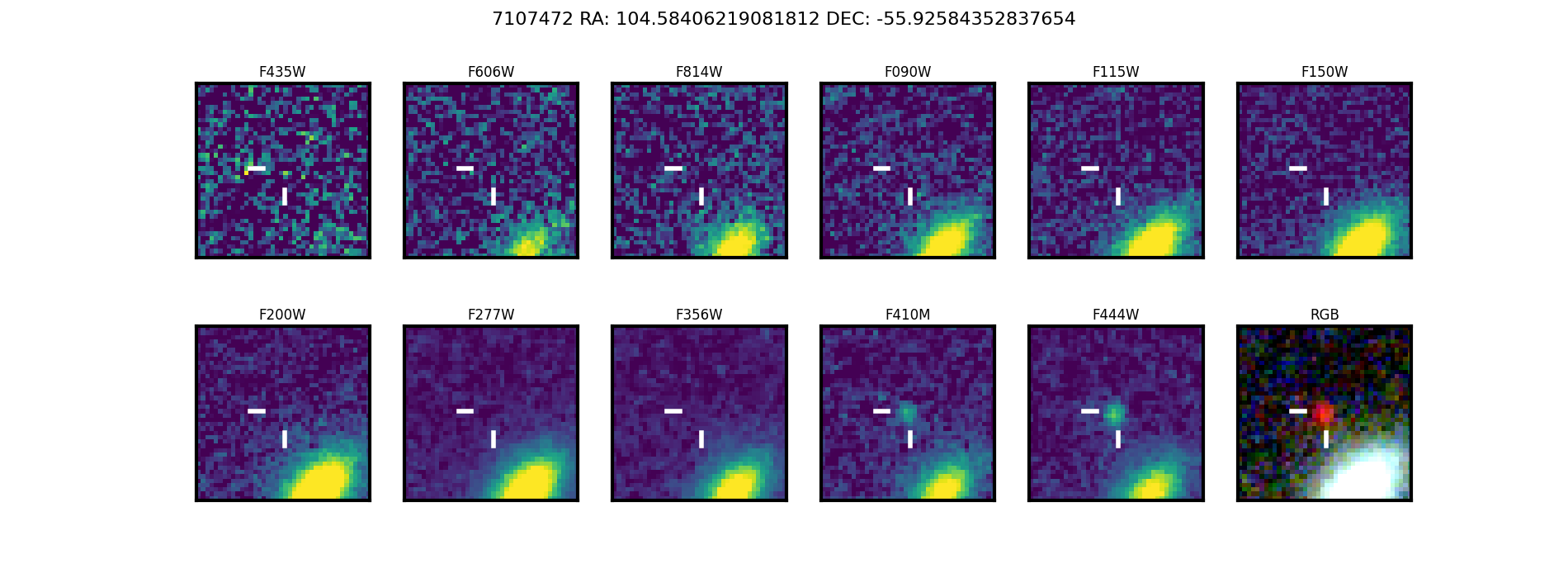}
\put(15,34){\textbf{Bullet-BD1}}
\end{overpic}
\begin{overpic}[width=1.0\textwidth]{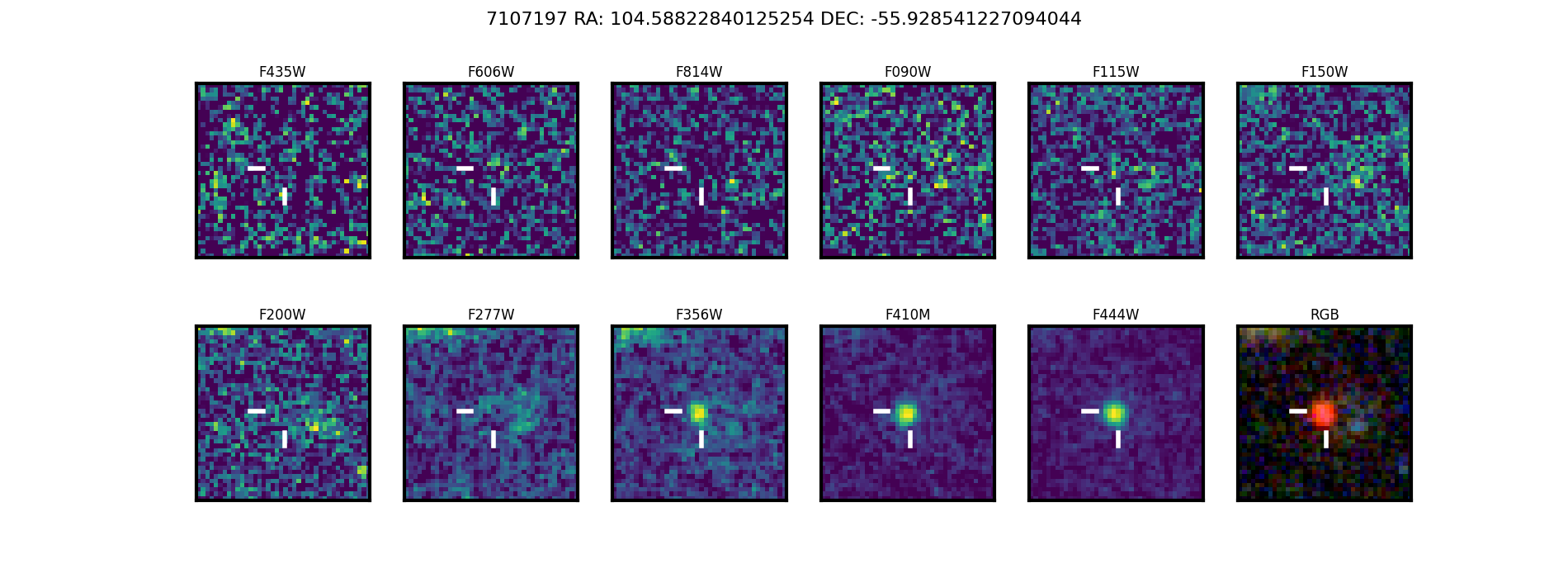}
\put(15,34){\textbf{Bullet-BD2}}
\end{overpic}
  \caption{$1\farcs6 \times1\farcs6$ cutouts of the F356W dropout (Bullet-BD1, ID 7107472, {\bf top}) and F277W dropout (Bullet-BD2, ID 7107197, {\bf bottom}) in the Bullet cluster data. Shown are different filters and an RGB image (B=F090W+F115W+F150W, G=F200W+F277W+F356W, and R=F356W+F410M+F444W). The images have not been convolved to the same PSF.} The white ticks mark the central position of the source (given by R.A., Decl. in Table \ref{tab:prop}).
    \label{fig:photometry}
 \end{figure*}

\section{Data}
 \defcitealias{canucsdr}{Sarrouh \& Asada et al. (2025)} 
\label{sec:data}
 Initial  {\jwst} NIRCam and NIRSpec observations of the Bullet Cluster were taken as part of the GO Program \#4598 	Silver Bullet for Dark Matter (co-PIs Brada\v{c}, Rihtar\v{s}i\v{c}, Sawicki, \citealp{rihtarsic26}).  The field was observed on 20 January 2025 (time stamp 2025-01-20 UTC09:39) with NIRCam imaging using filters F090W, F115W, F150W, F200W, F277W, F356W, F410M, and F444W with exposure times of $6.4\,\mbox{ks}$ each, reaching S/N between 5 and 10 for a $m_{\rm AB} = 29$ point source.  We used the FULLBOX dither pattern to cover the gap between the two NIRCam modules. We also utilized archival data of {\it HST}/ACS imaging from HST-GO-10200 (PI Jones), 10863 (PI Gonzalez), and 11099 (PI Brada\v{c}). NIRSpec Micro-Shutter Assembly (MSA) observations of targets selected from the NIRCam and ACS imaging were executed on 27 March 2025. The observations consisted of three separate MSA configurations with the low-resolution prism, using the standard 3 shutter nodding procedure with total integration time of $3.5\,\mbox{ks}$ per MSA configuration.  

 To assess variability and proper motion, we also used data from JWST-GO program \#6882 Vast Exploration for Nascent, Unexplored Sources (VENUS, co-PIs Fujimoto, Coe) that re-observed the Bullet Cluster $1.003$ years after the initial NIRCam campaign of \#4598 in the F150W, F210M, F300M, and F444W filters. Here we use the F444W observations with total integration time of $1.2\, \mbox{ks}$, taken on 2026-01-21 UTC18:27.

We use the CANUCS pipeline presented in more detail in \citetalias{canucsdr} to reduce the imaging data (including HST). Briefly, the raw data have been reduced using the STScI pipeline (stage 1). Stage 2 level of the STScI pipeline is accomplished using a combination of standard and custom steps, which mask imaging artefacts. For further steps, we use \texttt{grizli} code \citep{Brammer2019} that provides astrometric calibrations based on the Gaia Data Release 3 catalog, and shifts images using {\astrodrizzle}. The photometric zero-points are applied as described in \citet{grizliphot} and \citetalias{canucsdr}. 

 Bright cluster members, foreground galaxies and intracluster light are removed from the images, following the procedure outlined in \citet{martis24} and \citetalias{canucsdr}. For the photometry we follow the CANUCS pipeline and procedure as described in \citetalias{canucsdr}. We build a $\chi_{\rm mean}$-detection image \citep{Drlica-Wagner2018a} by combining all images before any PSF (point spread function) convolution. An empirical PSF in each filter was generated by combining stars in the field as described in \citetalias{canucsdr}. Given that our targets of interest are typically small, we use the aperture with $0.3\arcsec$ diameter to measure fluxes for colors. All fluxes are measured on images that have been convolved to the PSF of the F444W image to ensure the same regions contribute at all wavelengths. Flux uncertainties are determined from the distribution of sky values measured in identically-shaped apertures located in empty regions of the image.
 
 We searched in the first epoch of NIRCam data for all potential high redshift objects. Two objects, Bullet-BD1 (catalog ID 7107472) and Bullet-BD2 (7107197), were identified from these data as F356W and F277W dropouts, respectively. With detections in only a few filters (Fig.~\ref{fig:photometry}), the nature of these sources was uncertain, so they were prioritized for spectroscopic followup. It was noted at the time that the two objects are close together on the sky at just $12.8\arcsec$ apart and therefore may be physically associated. We give their properties in Table~\ref{tab:prop}. 
 
 \begin{figure}
\begin{overpic}[width=0.5\textwidth]{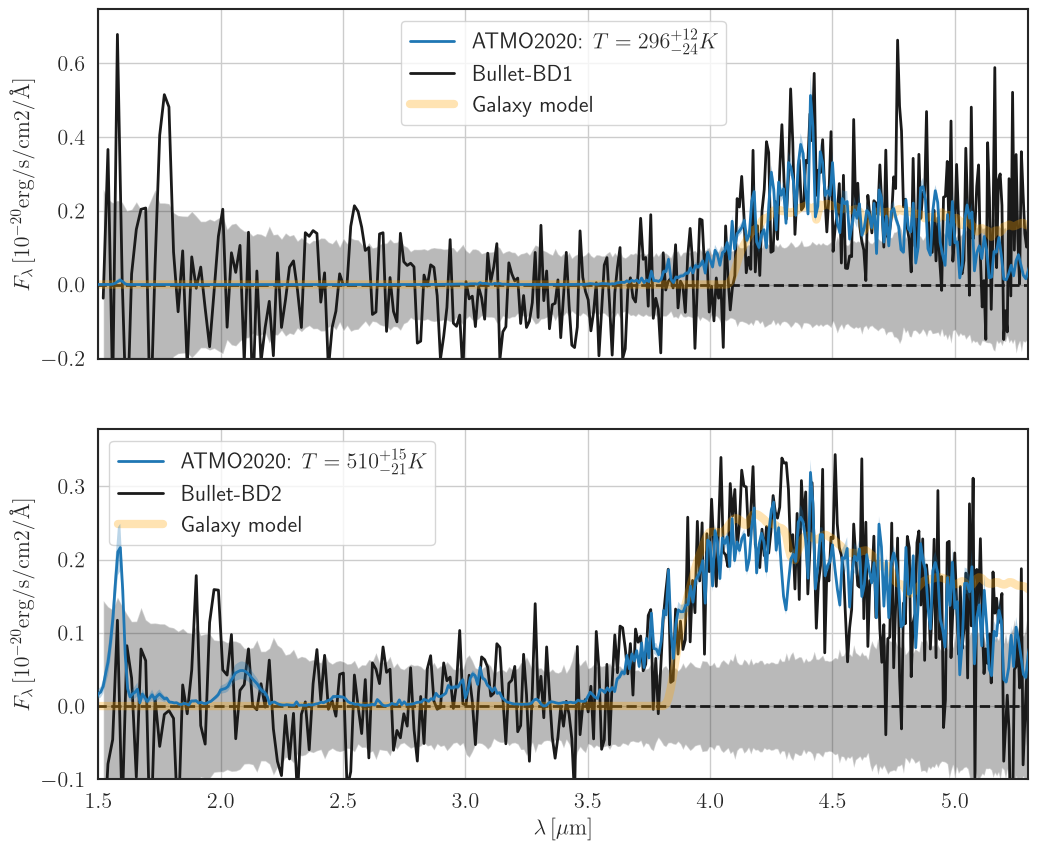}
     \put(20,65){\includegraphics[width=0.07\textwidth]{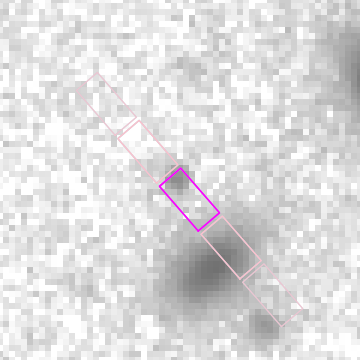}}
     \put(43,25){\includegraphics[width=0.07\textwidth]{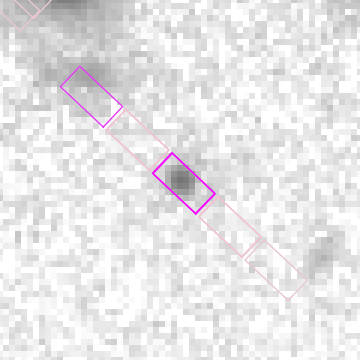}}
  \end{overpic}  
    \caption{NIRSpec PRISM Spectra of the F356W dropout (Bullet-BD1, 7107472, {\bf top}) and F277W dropout (Bullet-BD2, 7107197, {\bf bottom}) with slit placement insets. Also shown are flux uncertainties in gray. We show model fits using {\tt \bdmodelA} model \citep{atmopp} built into {\nifty} \citep{hainline25} in blue. In yellow, we show best-fit galaxy models fits (at $z=32.2\pm 0.5$ and $30.0\pm 0.1$) using {\bagpipes}. Brown dwarf templates fit the data marginally better ($\chi^2/{\rm DOF}=1.26$ vs. 1.27 for Bullet-BD1 and $\chi^2/{\rm DOF}=0.97$ vs. 1.23 for Bullet-BD2).}
    \label{fig:spectra}
 \end{figure}
 The objects were then followed up with NIRSpec (Fig.~\ref{fig:spectra}). Details of the NIRSpec processing are given in \citetalias{canucsdr} and  \citet{rihtarsic26}. Initial processing uses the STScI \texttt{jwst} stage 1 pipeline with custom snowball and 1/{\it f} noise correction. The \texttt{jwst} stage 2 pipeline is run up to the photometric calibration step followed by the \texttt{grizli} \citep{grizli23} and \texttt{msaexp}  \citep{msaexp} packages. The wavelength calibration uses a correction for the known intra-shutter offset along the dispersion direction. The spectral background is removed using the standard nodded background subtraction. One-dimensional spectra are extracted using a wavelength-dependent optimal extraction that accounts for the increase in PSF FWHM with wavelength. For one of the two objects (Bullet-BD1), the background was oversubtracted due to the bright neighboring galaxy in our slit (Fig.~\ref{fig:spectra} inset), and we apply a constant offset of $+0.012\,\mu\mbox{Jy}$, measured blueward of $3.5\mu\mbox{m}$ to correct for this. This is also consistent with photometry.

 \begin{figure*}
 \centering
 \includegraphics[width=0.99\textwidth]{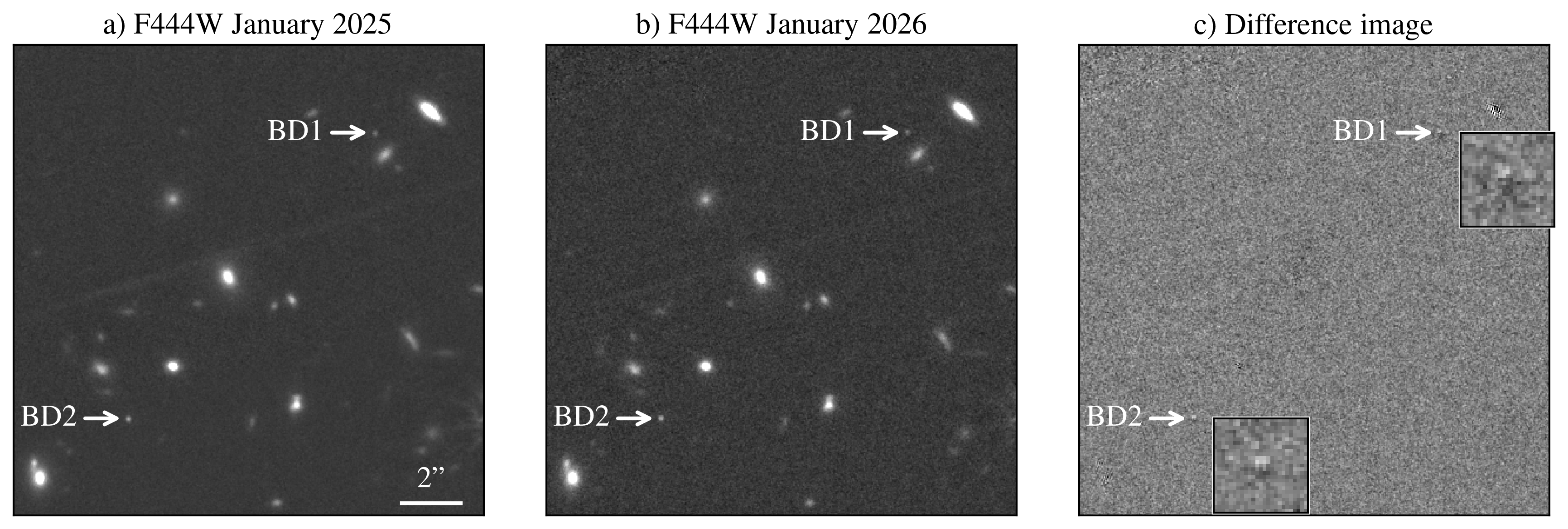}
  \includegraphics[width=0.85\textwidth]{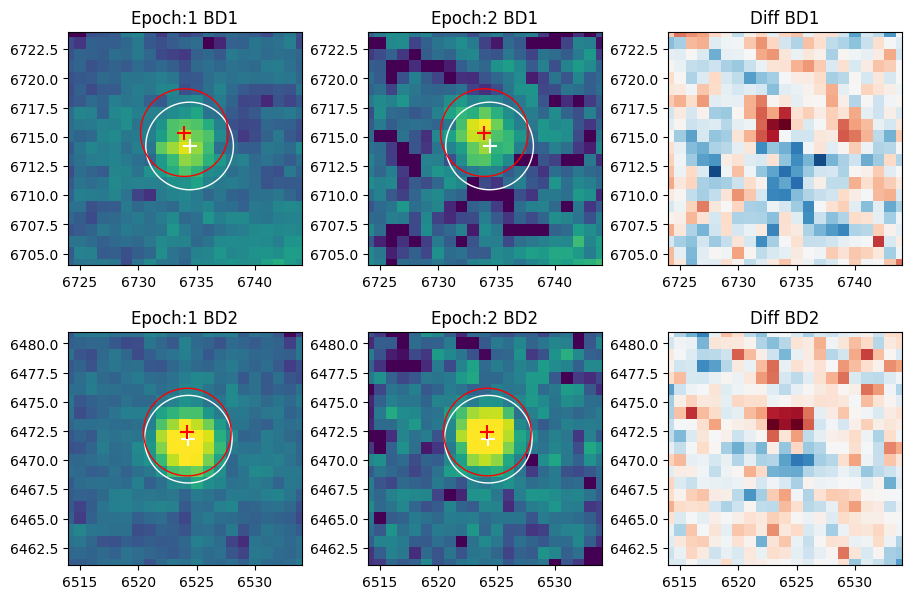}
  \caption{Detected proper motion of the two objects. {\bf Top:} Shown is a $16\arcsec \times 16\arcsec$ cutout of the field with Bullet-BD1 and Bullet-BD2 marked. Left column (a) shows the original F444W image taken by the JWST Silver Bullet collaboration (epoch 1), middle column (b) is the image taken $\sim 1$ year later by VENUS collaboration (epoch 2) and right panel (c) shows the difference image between epoch 2 and 1 with zoomed insets for the two objects of interest. {\bf Middle and Bottom:} Zoom in to the same images (for difference, red is positive, blue is negative). Both objects can be seen moving (white circular aperture is for epoch 1, red for epoch 2, crosses denote their centroids) in a similar direction ({\dir} measured N through E) with proper motion of {\pmo} for Bullet-BD1 and {\pmt} for Bullet-BD2.}
    \label{fig:pm}
 \end{figure*}

\begin{deluxetable}{lcc} 
\tabletypesize{\footnotesize}
\tablecolumns{5}
\tablewidth{0pt}
\tablecaption{Properties of Bullet-BD1 and Bullet-BD2\label{tab:prop}. All errors are given as $1\sigma$ except for upper-limits, which we quote as 2$\sigma$ for flux and 3$\sigma$ for sizes.}
\tablehead{\colhead{Parameter} & \colhead{Bullet-BD1 (7107472)} & \colhead{Bullet-BD2 (7107197)}}
\startdata
R.A. (deg) &  104.5840622 & 104.5882284\\
Decl. (deg) & $-55.9258435$& $-55.9285412$ \\
$mag_{\rm AB}(F444W)$ & $27.52\pm 0.04$ & $26.89\pm 0.01$\\ 
Size & $<0.071\arcsec$&  $<0.043\arcsec$ \tablenotemark{(a)} \\
\hline
 & \multirow{3}{*}{\raisebox{5ex}{{\bdmodelA} model}} & \\
$T_{\rm eff}$ & ${\toneA}$ & ${\ttwoA}$\\
Distance ($R=R_{\rm jup}$) & {\dteffoA} & {\dtefftA} \tablenotemark{(b)}\\ 
\hhline{---}
 & \multirow{3}{*}{\raisebox{5ex}{\bdmodelS}} & \\
$T_{\rm eff}$ & ${\toneS}$ & ${\ttwoS}$\\
Distance ($R=R_{\rm jup}$) & {\dteffoS} & {\dtefftS} \tablenotemark{(b)}\\ 
\hline
Proper motion (mas/yr) &{\pmno} & {\pmnt}\\
\hline
F200W (nJy) & $<6$& $<6$  \\
F277W (nJy) & $<6$& $<3$  \\
F356W (nJy) & $<6$& $15 \pm 1$  \\
F410M (nJy) & $34 \pm 3$& $90 \pm 3$  \\
F444W (nJy) & $51 \pm 2$& $92 \pm 2$  \\
F444W flux ratio (Epoch 2/ Epoch 1) & $0.98 \pm 0.07$ & $1.02 \pm 0.03$ \\
\enddata
\tablenotetext{(a)}{ Given are $3\sigma$ upper limits of the sizes.}
\tablenotetext{(b)}{To calculate distances from the spectral fit we assume the radius of brown dwarf of $R=R_{\rm jup}$. The errors on the distance do not include errors on $R$.}
\end{deluxetable}

\section{Data Analysis}
\label{sec:dataanalysis}

\subsection{Photometry and Variability}
Photometric fluxes and errors were determined using the aperture with $0.3\, \arcsec$ diameter on PSF-convolved images and corrected with a correction factor of $1/0.7$ to account for the flux losses in the aperture \citepalias{canucsdr}. In addition, we measure flux on the F444W VENUS image, with the position corrected for proper motion. To determine the proper motions of the objects, we first build a difference image to visualize the situation. The two dropouts clearly stand out as variable in position, whereas there are no visible asymmetric residuals around the other nearby objects (Fig.\ref{fig:pm}). We use the {\texttt centroid\_2dg} routine from {\photutils} to measure the centroids in both F444W background-subtracted images (the only filter with both epochs in which we detect objects). The proper motion is clearly detected in both objects, as shown in Fig.~\ref{fig:pm} and Table~\ref{tab:prop}. Uncertainties on proper motions are determined by inserting $300$ simulated sources (empirical PSFs from Section \ref{sec:data} with the same total fluxes as the two objects) into the F444W epoch 1 and 2 images and measuring the distribution of position offsets compared to the known input positions.

\subsection{Spectral Fitting}
Given the spectral shape of both objects, detected proper motion, and lack of variability (see Section~\ref{sec:results}), we conclude that both objects are Y dwarfs. For the brown dwarf models we use the fitting routine {\nifty} \citep{hainline25}. {\nifty} compares observed near-to-mid-infrared spectroscopy to standard brown dwarf models in a Bayesian framework. We use {\tt \bdmodelA} atmospheric model \citep{atmopp} that covers the range in effective temperature $250\,{\rm K}\leq T_{\rm eff}\leq1200\,{\rm K}$, surface gravity $2.5\leq \log g \leq 5.5 \,(\mbox{cgs})$, and solar scaled metallicity  $-1\leq [M/H] \leq 0.3$. Given that {\bdmodelA} reaches the lowest temperature among the available models, we use it as a fiducial model. For comparison, we also use {\bdmodelS} model \citep{elfowl25}, for which the ranges are $275\,{\rm K}\leq T_{\rm eff}\leq2400\,{\rm K}$, $3\leq \log g \leq 5 \,(\mbox{cgs})$, and $-1\leq [M/H] \leq 1$. We further fit spectra with standard galaxy models using {\bagpipes} \citep{carnall18} that include the damping wing prescription as described in \citet{asada25}.

\section{Results}
\label{sec:results}
\subsection{Imaging}
All the photometric flux measurements and errors are given in Table~\ref{tab:prop}. In addition, we also compare the F444W fluxes at the two epochs and do not detect significant variability in either object with flux ratios within $1\sigma$ error of 1.0.  We use {\galfit} \citep{peng11} to determine the sizes for both objects. Both objects are found to be unresolved with 3$\sigma$ upper limits on sizes of $<0.071\,\mbox{arcsec}$ for Bullet-BD1 and $<0.043\,\mbox{arcsec}$ for Bullet-BD2. 

The two epochs of NIRCam F444W imaging were separated by 1.003 years. From those images, we measure a proper motion of {\pmo} for Bullet-BD1 and {\pmt} for Bullet-BD2 (Fig.~\ref{fig:pm}). The significant detections of proper motion, at 6 and 8 $\sigma$, respectively, conclusively prove that these are galactic objects.The values are larger than average proper motion values of $\sim 3.06 \pm  0.67 \mbox{mas/yr}$ (dispersion $6.7\mbox{mas/yr}$) reported by \citet{windhorst11}, who measure proper motions of 103 faint ($16\mbox{mag}<AB<26\mbox{mag}$) stars in GOODS-S. They are comparable, however, to the proper motion detected for the brown dwarfs observed with JWST (e.g., \citealp{hainline24a, hainline24b, hainline25}). Both objects are found to be moving in a similar direction ({\dir} measured N through E).  

It is curious that the proper motions of Bullet-BD1 and Bullet-BD2 are not statistically consistent with each other, because the two BDs are separated by only $12.8\arcsec$ on the sky. We later comment on the implications of this measurement. 

%To determine the chance coincidence that the only two extremely cool Y dwarfs in the epoch 1 NIRCam imaging area would be so close together, we calculate the probability that an object is randomly located within $12.8\arcsec$ of another object, when only two such objects are detected in the imaged area. Note this calculation does not assume anything about the space density of Y dwarfs, as it only considers how likely it is that when there are two random objects in the field, they lie close together. This analysis takes into account the area covered at sufficient depth with NIRCam for detection and identification and the lack of detection of sources superimposed on existing sources. We take Bullet-BD2 as the reference position and ask how often one would find another object within $12.8\arcsec$ if only one other object were in the imaged area. We find only a 2\% probability that the close separation of these two objects occurs by chance.

\subsection{Spectra}

The results of fitting the NIRSpec spectra of both objects show that they are consistent with model spectra of brown dwarfs. While the S/N is low for Bullet-BD1, we still obtain a reasonable fit, giving an effective temperature of {\teffoA} for {\bdmodelA} and {\teffoS} for {\bdmodelS}. This is one of the lowest temperature brown dwarfs known spectroscopically with a best-fit temperature only slightly larger than the confirmed lowest temperature known brown dwarf, WISE\,J085510.83-071442.5 (Fig.~\ref{fig:ct}, \citealp{luhman2024}, our fit using {\bdmodelA} gives $T_{\rm eff}=263.0^{+0.2}_{-0.2}K$). In Fig.~\ref{fig:ct} we also show two other Y dwarfs from \citet{beiler24} WISEA\,J2354+02 with $T_{\rm eff}=359K$ and WISE\, J2209+27 $T_{\rm eff}=367K$). Bullet-BD2 is brighter and slightly less red than Bullet-BD1 in the wavelength range 3.5 to 4.5 $\mu$m. We find a best-fit temperature of {\tefftA} for {\bdmodelA} and {\tefftS} for {\bdmodelS} for Bullet-BD2.

To estimate the distances, {\nifty} uses the normalization
of each model to the spectra. Assuming the sizes of Bullet-BD1 and Bullet-BD2 are the same as that of Jupiter, we obtain distance estimates of {\dteffoA} and {\dteffoS} for Bullet-BD1 and {\dtefftA} and {\dtefftS} for Bullet-BD2 for {\bdmodelA} and {\bdmodelS} models, respectively (Table~\ref{tab:prop}). We caution, however, that there are known systematic errors in model-inferred temperature and radii from such fitting \citep{sanghi23}. We do not include these in the error budget, as they were not determined for the Y dwarfs.

The model distances are similar to those of other brown dwarfs found by JWST (e.g., \citealp{burgasser24,hainline25}). However, Bullet-BD1 is the lowest-temperature distant BD spectroscopically confirmed by JWST to date. If we use the distances derived from {\bdmodelA} and estimates from the proper motion (Table~\ref{tab:prop}), the tangential velocities of those objects would be $37^{+12}_{-8}\mbox{kms} ^{-1}$ for BD1 and $67\pm 15\mbox{kms}^{-1}$ for BD2.
The distances and velocities indicate that these BDs are likely members of the thick disk (e.g.~\citealp{dupuy12}).

Finally, we fit the spectra with standard galaxy templates using {\bagpipes}. The fits also include an additional damping wing component (see \citealp{asada25}). Reasonable fits, though marginally worse compared to the BD templates ($\chi^2/{\rm DOF}=1.26$ vs. 1.27 for Bullet-BD1 and $\chi^2/{\rm DOF}=0.97$ vs. 1.23 for Bullet-BD2 for brown dwarf/galaxy templates, respectively) are obtained (Fig.~\ref{fig:spectra}). However, the resulting galaxy templates  put these objects at extremely high redshifts ($z=32.2\pm 0.5$ and $z=30.0\pm 0.1$ , respectively) and extreme stellar masses for that redshift ($M^*\gtrsim 10^{10}M_{\odot}$, even after accounting for magnification of $\mu \sim 2$). Together with the proper motion data, we conclude that these objects are indeed brown dwarfs.

We also note that two other brown dwarfs (initially selected as high-redshift objects at $z\sim 7$) were targeted in our program. These objects with IDs 7103870 and 7104781 (the former is also presented in \citealp{tu25}) are higher temperature brown dwarfs ($>1000\mbox{K}$), and we see no significant proper motions, despite their clear brown dwarf spectral signatures. 

\section{Discussion} \label{sec:discussion}

\subsection{Nature of the two brown dwarfs}

Both brown dwarfs are well fit by extremely low temperature models, indicating they should be classified as Y dwarfs. These objects are both likely to be very old and low mass. To give a representative example, the inferred masses of BDs with $T_{\rm eff} = 300$\,K and $T_{\rm eff} = 400$\,K using the ATMO2020 evolutionary models \citep{phillips2020} are $\sim12$\,\Mjup\, and $\sim22$\,\Mjup\, at 10\,Gyr, or $\sim7$\, \Mjup\, and $\sim12$\,\Mjup\, at 3\,Gyr. Their estimated distances of 200 to 500\,pc and tangential velocities mean they could be part of the thick disk. They are also among the coldest BDs detected spectroscopically with JWST (e.g., \citealp{burgasser24, hainline24b, tu25, morrissey26}).

The finding of two such objects within the same field, and with such a small sky separation, is intriguing and raises the question of whether the two brown dwarfs are physically associated. However, several arguments suggest that they are not.
\begin{itemize}
    \item{The proper motions differ by $25\pm9$\,\mbox{mas/yr} (i.e. at the 3$\sigma$ level) with Bullet-BD2 having only half the proper motion of Bullet-BD1. That is irreconcilable with the much smaller orbital motion expected for a bound system, whose maximum amplitude is estimated using Kepler's third law at $\lesssim0.1$\,\mbox{mas/yr}, assuming a pair with total mass of 50\,\Mjup\, on a circular, face-on orbit.}
    \item{If placed at a common distance of $\sim 300\,\mbox{pc}$, the physical separation would be $4000\,\mbox{AU}$ or 0.02\,pc. This is an order of magnitude larger than the currently known largest separation of binary brown dwarfs: i.e. the pair CWISE J165141.67+695306.6 + CWISE J165143.63+695259.4 ($\sim700\,\mbox{AU}$), the pair CWISE J022454.10+152633.8 + CWISE J022454.80+152629.5 ($\sim430\,\mbox{AU}$, \citealp{rothermich2024}) and CWISE J014611.20–050850.0AB ($\lesssim 150\,\mbox{AU}$,  \citealp{softich2022}). Considering disruption by small perturbations due to passing stars at moderate distances, the average lifetime of wide binary systems in the Galaxy has been studied by \cite{weinberg1987}. The formalism involves the galactic stellar density and average perturber mass among other parameters, but has been reduced to depend on the total mass, $M_{tot}$, and age in Gyr, $t^{\star}$, of the binary system (see equation 18 of \citealp{dhital2010}) such that statistically, the widest binary surviving at separation $a$ is given by:
    \begin{equation}
    a \simeq 1.212 \frac{M_{tot}}{t^{\star}} \,{\rm pc.}
    \end{equation}
    It can be shown that the masses inferred from ATMO2020 evolutionary models \citep{phillips2020} of our 2 BDs correspond to ages always larger than the galactic disruption lifetime. Using the mass examples given at the beginning of this section yields binary lifetimes of 1.2~Gyr or 2.1~Gyr, which are smaller than the evolutionary age (3~Gyr or 10~Gyr), suggesting that our 2 BDs could not have survived to their current age without being disrupted.}

    \item{The binding energy of a binary system depends on the masses and separation of its members (see equation 3 of \citealt{rothermich2024}). No binary system is known with binding energy less than $3\times10^{39}$ ergs (Figure 11 of \citealt{rothermich2024}). Our BDs would have even lower binding energy, between $4\times10^{38}$ ergs and $3\times10^{39}$ ergs, for the age and mass examples given above.}
    
    \item{Bullet-BD2 is brighter and hotter than Bullet-BD1 with a flux difference of factor $\approx 2$ and temperature difference of factor $1.7^{+0.2}_{-0.1}$ (using \bdmodelA). If the two brown dwarfs are at the same distance, one can determine their relative radii from their luminosities and temperatures. Assuming luminosity is proportional to radius$^2\times T_{\rm eff}^4$, Bullet-BD2 would have a $35\% \pm 10\%$ smaller radius than Bullet-BD1 (note, however, that the errors on temperature do not include e.g., model-selection systematics). This is very difficult to reconcile with the ATMO2020 evolutionary model predictions, where a difference of less than 10\% is predicted for 2 BDs sharing the same age, and with the masses given above as representative examples.}
    
    \item Given the low $T_{\rm eff}$, they are likely old BDs, so must have wandered in the Galaxy for more than 100 Myr (inferred from evolutionary model tracks that reach $T_{\rm eff}$). They are thus unlikely to originate from the same birth cloud. 
\end{itemize}

 \begin{figure}
  \includegraphics[width=0.5\textwidth]{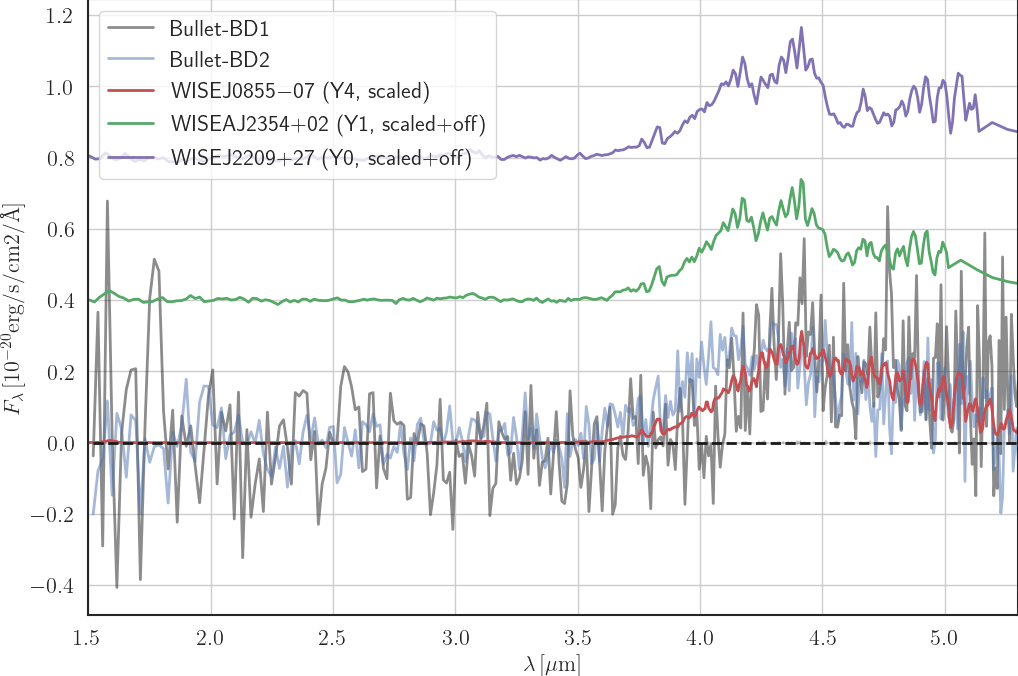}
  \includegraphics[width=0.5\textwidth]{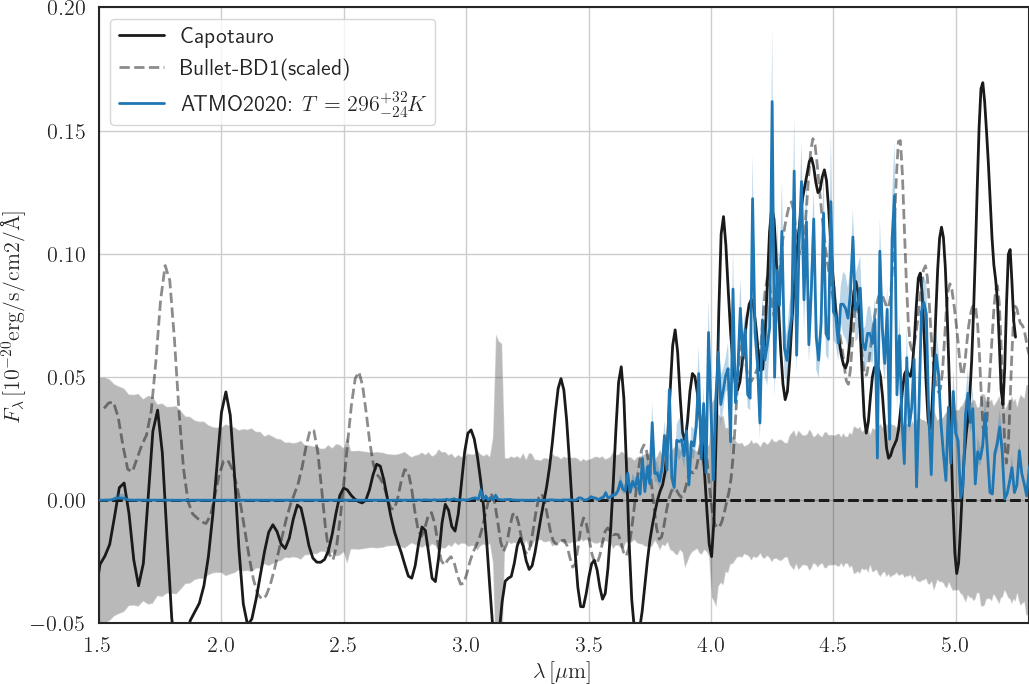}
  \caption{ {\bf (top)} Comparison of Bullet-BD1 (black), Bullet-BD2 (blue), and WISE\,J085510.83-071442.5  spectrum (scaled to match the flux between 1.5 and 5.3$\mu\mbox{m}$) from \citet{luhman2024} (red), indicating BD1 and WISE0855 have similar characteristics. We additionally plot low temperature (Y dwarfs) WISEA\,J2354+02 with $T_{\rm eff}=359K$ and WISE\, J2209+27 $T_{\rm eff}=367K$ from \citet{beiler24} (scaled and offset for clarity). {\bf (bottom)} Comparison of Capotauro spectrum (black line, grey-shade noise) with Bullet-BD1 spectrum (scaled by a factor 2.67 to match the fluxes in the wavelength range of 4.2 to $4.6\mu\mbox{m}$). Also shown is the best-fit brown dwarf model spectrum, derived with NIFTY and {\bdmodelA}, in blue. Both observed spectra were smoothed with a Gaussian ($\sigma = 2$) for a better visual comparison.}
    \label{fig:ct}
 \end{figure}

Considering these factors, we conclude that the two brown dwarfs are unlikely to be physically associated and are likely at significantly different distances, with the hotter and brighter brown dwarf farther away.

\subsection{Expected brown dwarfs number count}

Assuming the known local density is representative of the whole Galaxy, a rough estimate of the expected number count of brown dwarfs in this survey can be established. We adopt the standard model for the star density distribution, $\rho$, in the Galaxy, for both the thin and thick disks, a double exponential disk \citep[eq. 19]{juric_milky_2008}:
\begin{equation}
\rho(R,z)=
\rho_{\odot}
\exp\!\left[-\frac{R-R_\odot}{L}\right]
\exp\!\left[-\frac{|z|-|z_\odot|}{H}\right]
\end{equation}
where $R$ and $z$ are distances expressed in cylindrical coordinates, $R_\odot=8000$\,pc, $z_\odot=25$\,pc are the Sun position, $L$ is the radial scale of the Galaxy and $H$ the scale height. Our calculations adopt values of $L=2600$\,pc and $H=300$\,pc for the thin disk, $L=3600$\,pc and $H=900$\,pc for the thick disk. Also, the $\rho_{\odot}$ for the thick disk has a value 0.12$\times$ that of the thin disk \citep[table 10]{juric_milky_2008}.

The number count of brown dwarfs, $n_{\rm BD}$, is calculated as
\begin{equation}
    n_{\rm BD} = \rho_{\odot}  \Omega  \int_0^{d_{\rm max}} \rho(s) s^2 ds
\end{equation}
where $\Omega$ is the survey area in steradians, $\rho_{\odot}$ is the local density of brown dwarfs per $\mbox{pc}^3$, $s$ is the distance in pc and the integral represents the effective volume normalized to the local density, integrated up to the survey distance limit, $d_{\rm max}$, set by the combination of the survey limiting magnitude and the brown dwarf absolute magnitudes.

The 20 pc local neighborhood sample of brown dwarfs is well characterized \citep[Table 15]{kirkpatrick21} with a density of Y dwarfs in the 300-600K regime of $\rho_{\odot,\rm Y} \geq 7.1\times10^{-3} pc^{-3}$. Adopting a survey area of 15 sq. arcmin, a limiting magnitude of F444W = 28.5 (needed for the photometric selection and spectroscopic follow-up, nominal 3$\sigma$ survey depth is 29.8), the Y-type BDs absolute Vega magnitude of M$_{\rm F444W} = 16.0$ \citep[Fig. 17]{kirkpatrick21}, results in a survey limiting distance of $d_{\rm max} = 724$\,pc and in an expected number count of $\geq0.7$ Y-dwarfs in this survey compared to the 2 discovered.

The finding of two, seemingly unassociated, cool Y dwarfs in this single extragalactic field implies a higher density of such objects in the Bullet cluster field than in JADES survey. Only a handful of similar objects have been reported so far. In the two JADES fields that cover an area 21 times larger than the Bullet cluster field, \cite{hainline25} report only 3 Y dwarfs with $T_{\rm eff}\leq 400\,K$ with similar depth in the medium survey (80\% of the fields) and 5-7 times longer exposure times in the deep and deepest surveys compared to our data. The coolest of these, JADES-GS-BD-5, has $T_{\rm eff} \approx 320\,K$ (estimated from photometry), which is similar to Bullet-BD1. Furthermore, 10 candidates have fits consistent with Y dwarf temperatures in the JADES fields. The space density of Y dwarfs in the Bullet cluster field is $\sim$ 4$\times$ higher than in high-galactic latitude extragalactic deep fields ($b=+59\deg, 55\deg$, and $-54\deg$, respectively). Our calculations also show that in a survey at the same depth and surface area as the Bullet Cluster data, but at higher latitudes, number counts of brown dwarfs would drop by a factor of $\gtrsim 2$.

We also note that some other high-redshift galaxy searches are conducted at relatively low galactic latitudes and caution that these searches are likely to misidentify extremely red Y dwarfs as high-z candidates visible only in the reddest NIRCam filters. At high galactic latitudes, their occurrence might be less likely, though not impossible. Also, the calculations indicate that the expected number count of all brown dwarfs (150-2250 K) should be a factor of $\sim3$ larger than that of Y-dwarfs alone. We have not carried out a comprehensive search for all brown dwarfs in our survey.

\subsection{Implications for extremely high-redshift galaxy searches}

The observations thus indicate that a non-negligible fraction of unresolved dropouts could be mistaken for brown dwarfs. The NIRSpec spectrum of Bullet-BD1 is also very similar to that for the Capotauro object in the CEERS field (Fig.~\ref{fig:ct}, \citealp{gandolfi25}). The temperature of their brown dwarf template fit is $\approx 300\,K$. Our template fit results in a similar temperature  (Fig.~\ref{fig:ct}), which is also similar to Bullet-BD1. Furthermore, our detected proper motion is below the limit achievable with the current Capotauro data ($\lesssim 137\,\mbox{mas/yr}$, \citealp{gandolfi25}). The main difference is, however, that Capotauro shows variability. We note, however, that both proper motion and variability were assessed by comparing NIRCam and NIRSpec data, so they are not as accurate as our data, which use two epochs of NIRCam data in the same filter. It is thus possible that Capotauro is also a Y dwarf (see also the discussion in \citealt{hainline25}). We expect the situation to be clarified further in mid 2026 with scheduled \jwst\ NIRCam medium-band imaging, which may determine the proper motion and possibly confirm the $4.4\mu$m peak in flux apparent in the rather noisy NIRSpec spectrum.

\section{Conclusions} \label{sec:conclusions}

We have investigated two extremely red objects from NIRCam imaging of the Bullet cluster. Using only single-epoch imaging data and no spectroscopy, such sources could be easily interpreted as reddened extreme emission-line galaxies at $z\sim8$ or Lyman-break galaxies at $z>20$. However, NIRSpec follow-up showed that the two objects show spectra characteristic of Y dwarfs. Fitting the spectra using {\nifty} reveals low temperatures {\tonerange} for Bullet-BD1 and {\ttworange} for Bullet-BD2  (using {\bdmodelA} and {\bdmodelS}   models). Furthermore, additional imaging, taken $\sim 1$ year after the original data, clearly reveals proper motion of both objects, {\pmo} and {\pmt}, confirming them as brown dwarfs. The two objects are in close proximity on the sky, but several lines of evidence point towards them not being physically associated. We determine a relatively high sky density of 0.13Y dwarfs per arcmin$^2$, likely the result of observing at a relatively low galactic latitude ($b=-21\deg$). At high galactic latitudes, their occurrence might be less likely, though not impossible.

\section*{Acknowledgements}
The authors would like to thank Giovanni Gandolfi for kindly providing the spectral data for Capotauro. MB, JJ, GR, NM, AH, and VM acknowledge support from the ERC Grant FIRSTLIGHT \# 101053208, Slovenian national research agency ARIS through grants N1-0238 and P1-0188, and ESA PRODEX Experiment Arrangements No. 4000146646 and 4000149972. JMD acknowledges the support of projects PID2022-138896NB-C51 (MCIU/AEI/MINECO/FEDER, UE) Ministerio de Ciencia, Investigaci\'on. KI acknowledges support from the National Natural Science Foundation of China (12573015, 1251101148, 12233001, 12473037), the Beijing Natural Science Foundation (IS25003), and the China Manned Space Program (CMS-CSST-2025-A09). LA ackowledges support from the Canadian Space Agency (CSA) grant 22EXPJWST and MS and YA from CSA grant 24JWGO3A12. DM acknowledges generous support from the Leonard and Jane Holmes Bernstein Professorship in Evolutionary Science. Support for program JWST-GO-06882, provided through a grant from the STScI under NASA contract NAS5-03127, is acknowledged. EV acknowledges financial support through grants INAF GO 2024 "Mapping Star Cluster Feedback in a Galaxy 450 Myr after the Big Bang" and by the European Union – NextGenerationEU within PRIN 2022 project n.20229YBSAN - Globular clusters in cosmological simulations and lensed fields: from their birth to the present epoch. RAW acknowledges support from NASA JWST Interdisciplinary Scientist grants NAG5-12460, NNX14AN10G and 80NSSC18K0200 from GSFC. BL is supported by the international Gemini Observatory, a program of NSF NOIRLab, which is managed by the Association of Universities for Research in Astronomy (AURA) under a cooperative agreement with the U.S. National Science Foundation, on behalf of the Gemini partnership of Argentina, Brazil, Canada, Chile, the Republic of Korea, and the United States of America. SF acknowledges support from the Dunlap Institute, funded through an endowment established by the David Dunlap family and the University of Toronto. 
We acknowledge the support of the Canadian Space Agency (CSA) [25JWGO4A18]. Support for SWR was provided by the Chandra X-ray Observatory Center, which is operated by the Smithsonian Astrophysical Observatory for and on behalf of
NASA under contract NAS8-03060.

This research was also enabled by grant 18JWST-GTO1 from the Canadian Space Agency and funding from the Natural Sciences and Engineering Research Council of Canada grants RGPIN-2020-06023 and RGPAS-2020-00065 to MS . This research used the Canadian Advanced Network For Astronomy Research (CANFAR) operated in partnership by the Canadian Astronomy Data Centre and The Digital Research Alliance of Canada with support from the National Research Council of Canada the Canadian Space Agency, CANARIE and the Canadian Foundation for Innovation. The data were obtained from the Mikulski Archive for Space Telescopes at the Space Telescope Science Institute, which is operated by the Association of Universities for Research in Astronomy, Inc., under NASA contract NAS 5-03127 for JWST. These observations are associated with program \#4598.  Support for program \#4598 was provided by NASA through a grant JWST-GO-4598 from the Space Telescope Science Institute, which is operated by the Association of Universities for Research in Astronomy, Inc., under NASA contract NAS 5-03127.

\section*{Data Availability}
All the {\it JWST} data used in this paper can be found in MAST: \dataset[10.17909/tmvk-zd12]{http://dx.doi.org/10.17909/tmvk-zd12}.
Data behind Fig.~\ref{fig:spectra} are provided in machine-readable format by the journal.

\facilities{HST (ACS,WFC3), JWST (NIRCam, NIRSpec)}

\bibliographystyle{aasjournal}
\bibliography{bibliogr_cv,bibliogr_highz, canucs_dr1, bibliogr_bd}

\begin{thebibliography}{}
\expandafter\ifx\csname natexlab\endcsname\relax\def\natexlab#1{#1}\fi
\providecommand{\url}[1]{\href{#1}{#1}}
\providecommand{\dodoi}[1]{doi:~\href{http://doi.org/#1}{\nolinkurl{#1}}}
\providecommand{\doeprint}[1]{\href{http://ascl.net/#1}{\nolinkurl{http://ascl.net/#1}}}
\providecommand{\doarXiv}[1]{\href{https://arxiv.org/abs/#1}{\nolinkurl{https://arxiv.org/abs/#1}}}

\bibitem[{{Adamo} {et~al.}(2025){Adamo}, {Atek}, {Bagley}, {Ba{\~n}ados}, {Barrow}, {Berg}, {Bezanson}, {Brada{\v{c}}}, {Brammer}, {Carnall}, {Chisholm}, {Coe}, {Dayal}, {Eisenstein}, {Eldridge}, {Ferrara}, {Fujimoto}, {Graaff}, {Habouzit}, {Hutchison}, {Kartaltepe}, {Kassin}, {Kriek}, {Labb{\'e}}, {Maiolino}, {Marques-Chaves}, {Maseda}, {Mason}, {Matthee}, {McQuinn}, {Meynet}, {Naidu}, {Oesch}, {Pentericci}, {P{\'e}rez-Gonz{\'a}lez}, {Rigby}, {Roberts-Borsani}, {Schaerer}, {Shapley}, {Stark}, {Stiavelli}, {Strom}, {Vanzella}, {Wang}, {Wilkins}, {Williams}, {Willott}, {Wylezalek}, \& {Nota}}]{adamo25}
{Adamo}, A., {Atek}, H., {Bagley}, M.~B., {et~al.} 2025, Nature Astronomy, 9, 1134, \dodoi{10.1038/s41550-025-02624-5}

\bibitem[{{Asada} {et~al.}(2025){Asada}, {Desprez}, {Willott}, {Sawicki}, {Brada{\v{c}}}, {Brammer}, {Dubath}, {Iyer}, {Martis}, {Muzzin}, {Noirot}, {Paltani}, {Sarrouh}, {Harshan}, \& {Markov}}]{asada25}
{Asada}, Y., {Desprez}, G., {Willott}, C.~J., {et~al.} 2025, \apjl, 983, L2, \dodoi{10.3847/2041-8213/adc388}

\bibitem[{{Asada} {et~al.}(2026){Asada}, {Willott}, {Muzzin}, {Brada{\v{c}}}, {Brammer}, {Desprez}, {Iyer}, {Marchesini}, {Martis}, {Noirot}, {Sarrouh}, {Sawicki}, {Withers}, {Fujimoto}, {Felicioni}, {Goovaerts}, {Jude{\v{z}}}, {Jagga}, {Merchant}, {M{\'e}rida}, \& {Robbins}}]{asada26}
{Asada}, Y., {Willott}, C.~J., {Muzzin}, A., {et~al.} 2026, \apj, 996, 115, \dodoi{10.3847/1538-4357/ae1f8d}

\bibitem[{{Beiler} {et~al.}(2024){Beiler}, {Cushing}, {Kirkpatrick}, {Schneider}, {Mukherjee}, {Marley}, {Marocco}, \& {Smart}}]{beiler24}
{Beiler}, S.~A., {Cushing}, M.~C., {Kirkpatrick}, J.~D., {et~al.} 2024, \apj, 973, 107, \dodoi{10.3847/1538-4357/ad6301}

\bibitem[{{Brammer}(2019)}]{Brammer2019}
{Brammer}, G. 2019, {Grizli: Grism redshift and line analysis software}, Astrophysics Source Code Library, record ascl:1905.001

\bibitem[{Brammer(2022)}]{grizliphot}
Brammer, G. 2022, {Preliminary updates to the NIRCam photometric calibration},  Zenodo, \dodoi{10.5281/zenodo.7143382}

\bibitem[{{Brammer}(2023{\natexlab{a}})}]{grizli23}
{Brammer}, G. 2023{\natexlab{a}}, {grizli}, 1.9.6, Zenodo,  Zenodo, \dodoi{10.5281/zenodo.1146904}

\bibitem[{{Brammer}(2023{\natexlab{b}})}]{msaexp}
---. 2023{\natexlab{b}}, {msaexp: NIRSpec analyis tools}, 0.6.17,  Zenodo, \dodoi{10.5281/zenodo.8319596}

\bibitem[{{Burgasser} {et~al.}(2024){Burgasser}, {Bezanson}, {Labbe}, {Brammer}, {Cutler}, {Furtak}, {Greene}, {Gerasimov}, {Leja}, {Pan}, {Price}, {Wang}, {Weaver}, {Whitaker}, {Fujimoto}, {Kokorev}, {Dayal}, {Nanayakkara}, {Williams}, {Marchesini}, {Zitrin}, \& {van Dokkum}}]{burgasser24}
{Burgasser}, A.~J., {Bezanson}, R., {Labbe}, I., {et~al.} 2024, \apj, 962, 177, \dodoi{10.3847/1538-4357/ad206f}

\bibitem[{{Carnall} {et~al.}(2018){Carnall}, {McLure}, {Dunlop}, \& {Dav{\'e}}}]{carnall18}
{Carnall}, A.~C., {McLure}, R.~J., {Dunlop}, J.~S., \& {Dav{\'e}}, R. 2018, \mnras, 480, 4379, \dodoi{10.1093/mnras/sty2169}

\bibitem[{{Carniani} {et~al.}(2024){Carniani}, {Hainline}, {D'Eugenio}, {Eisenstein}, {Jakobsen}, {Witstok}, {Johnson}, {Chevallard}, {Maiolino}, {Helton}, {Willott}, {Robertson}, {Alberts}, {Arribas}, {Baker}, {Bhatawdekar}, {Boyett}, {Bunker}, {Cameron}, {Cargile}, {Charlot}, {Curti}, {Curtis-Lake}, {Egami}, {Giardino}, {Isaak}, {Ji}, {Jones}, {Kumari}, {Maseda}, {Parlanti}, {P{\'e}rez-Gonz{\'a}lez}, {Rawle}, {Rieke}, {Rieke}, {Del Pino}, {Saxena}, {Scholtz}, {Smit}, {Sun}, {Tacchella}, {{\"U}bler}, {Venturi}, {Williams}, \& {Willmer}}]{carniani24}
{Carniani}, S., {Hainline}, K., {D'Eugenio}, F., {et~al.} 2024, \nat, 633, 318, \dodoi{10.1038/s41586-024-07860-9}

\bibitem[{{Castellano} {et~al.}(2025){Castellano}, {Fontana}, {Merlin}, {Santini}, {Napolitano}, {Menci}, {P{\'e}rez-Gonz{\'a}lez}, {Calabr{\`o}}, {Paris}, {Pentericci}, {Zavala}, {Dickinson}, {Finkelstein}, {Treu}, {Amorin}, {Arrabal Haro}, {Bergamini}, {Bisigello}, {Catone}, {Daddi}, {Dayal}, {Dekel}, {Ferrara}, {Fortuni}, {Gandolfi}, {Giavalisco}, {Grillo}, {Guida}, {Hathi}, {Holwerda}, {Koekemoer}, {Kokorev}, {Li}, {Llerena}, {Lucas}, {Mascia}, {Metha}, {Morishita}, {Nanayakkara}, {Pacucci}, {Roberts-Borsani}, {Rodighiero}, {Rosati}, {Salazar}, {Schneider}, {Somerville}, {Taylor}, {Trenti}, {Trinca}, {Wang}, {Watson}, {Yang}, \& {Yung}}]{castellano25}
{Castellano}, M., {Fontana}, A., {Merlin}, E., {et~al.} 2025, \aap, 704, A158, \dodoi{10.1051/0004-6361/202555082}

\bibitem[{{Dhital} {et~al.}(2010){Dhital}, {West}, {Stassun}, \& {Bochanski}}]{dhital2010}
{Dhital}, S., {West}, A.~A., {Stassun}, K.~G., \& {Bochanski}, J.~J. 2010, \aj, 139, 2566, \dodoi{10.1088/0004-6256/139/6/2566}

\bibitem[{Drlica-Wagner {et~al.}(2018)Drlica-Wagner, Sevilla-Noarbe, Rykoff, Gruendl, Yanny, Tucker, Hoyle, Rosell, Bernstein, Bechtol, Becker, Benoit-L{\'{e}}vy, Bertin, Kind, Davis, de~Vicente, Diehl, Gruen, Hartley, Leistedt, Li, Marshall, Neilsen, Rau, Sheldon, Smith, Troxel, Wyatt, Zhang, Abbott, Abdalla, Allam, Banerji, Brooks, Buckley-Geer, Burke, Capozzi, Carretero, Cunha, D'Andrea, da~Costa, DePoy, Desai, Dietrich, Doel, Evrard, Neto, Flaugher, Fosalba, Frieman, Garc{\'{i}}a-Bellido, Gerdes, Giannantonio, Gschwend, Gutierrez, Honscheid, James, Jeltema, Kuehn, Kuhlmann, Kuropatkin, Lahav, Lima, Lin, Maia, Martini, McMahon, Melchior, Menanteau, Miquel, Nichol, Ogando, Plazas, Romer, Roodman, Sanchez, Scarpine, Schindler, Schubnell, Smith, Smith, Soares-Santos, Sobreira, Suchyta, Tarle, Vikram, Walker, Wechsler, \& Zuntz}]{Drlica-Wagner2018a}
Drlica-Wagner, A., Sevilla-Noarbe, I., Rykoff, E.~S., {et~al.} 2018, Astrophys. J. Suppl. Ser., 235, 33, \dodoi{10.3847/1538-4365/aab4f5}

\bibitem[{{Dupuy} \& {Liu}(2012)}]{dupuy12}
{Dupuy}, T.~J., \& {Liu}, M.~C. 2012, \apjs, 201, 19, \dodoi{10.1088/0067-0049/201/2/19}

\bibitem[{{Ferrara} {et~al.}(2026){Ferrara}, {Carniani}, {Morishita}, \& {Stiavelli}}]{ferrara26}
{Ferrara}, A., {Carniani}, S., {Morishita}, T., \& {Stiavelli}, M. 2026, arXiv e-prints, arXiv:2601.07374, \dodoi{10.48550/arXiv.2601.07374}

\bibitem[{{Gandolfi} {et~al.}(2025){Gandolfi}, {Rodighiero}, {Castellano}, {Fontana}, {Santini}, {Dickinson}, {Finkelstein}, {Catone}, {Calabr{\`o}}, {Merlin}, {Pentericci}, {Bisigello}, {Grazian}, {Napolitano}, {Vulcani}, {Taylor}, {Arrabal Haro}, {Kirkpatrick}, {Backhaus}, {Holwerda}, {Giulietti}, {Bianchetti}, {Cassata}, {Cleri}, {Daddi}, {Ferguson}, {Girardi}, {Hirschmann}, {Koekemoer}, {Lapi}, {Pacucci}, {P{\'e}rez-Gonz{\'a}lez}, {de la Vega}, {Vietri}, {Wilkins}, {Yung}, {Bagley}, {Bhatawdekar}, {Kartaltepe}, {Papovich}, \& {Pirzkal}}]{gandolfi25}
{Gandolfi}, G., {Rodighiero}, G., {Castellano}, M., {et~al.} 2025, arXiv e-prints, arXiv:2509.01664, \dodoi{10.48550/arXiv.2509.01664}

\bibitem[{{Gandolfi} {et~al.}(2026){Gandolfi}, {Rodighiero}, {Bisigello}, {Grazian}, {Finkelstein}, {Dickinson}, {Castellano}, {Merlin}, {Calabr{\`o}}, {Papovich}, {Bianchetti}, {Ba{\~n}ados}, {Benotto}, {Catone}, {Buitrago}, {Daddi}, {Girardi}, {Giulietti}, {Hirschmann}, {Holwerda}, {Arrabal Haro}, {Lapi}, {Lucas}, {Lyu}, {Massardi}, {Pacucci}, {P{\'e}rez-Gonz{\'a}lez}, {Ronconi}, {Tarrasse}, {Wilkins}, {Vulcani}, {Yung}, {Zavala}, {Backhaus}, {Bagley}, {Buat}, {Burgarella}, {Kartaltepe}, {Khusanova}, {Kirkpatrick}, {Kocevski}, {Koekemoer}, {Lambrides}, {Pirzkal}, \& {Yang}}]{gandolfi26}
{Gandolfi}, G., {Rodighiero}, G., {Bisigello}, L., {et~al.} 2026, \aap, 708, A195, \dodoi{10.1051/0004-6361/202554009}

\bibitem[{{Hainline} {et~al.}(2024{\natexlab{a}}){Hainline}, {Helton}, {Johnson}, {Sun}, {Topping}, {Leisenring}, {Baker}, {Eisenstein}, {Hausen}, {Hviding}, {Lyu}, {Robertson}, {Tacchella}, {Williams}, {Willmer}, \& {Roellig}}]{hainline24a}
{Hainline}, K.~N., {Helton}, J.~M., {Johnson}, B.~D., {et~al.} 2024{\natexlab{a}}, \apj, 964, 66, \dodoi{10.3847/1538-4357/ad20d1}

\bibitem[{{Hainline} {et~al.}(2024{\natexlab{b}}){Hainline}, {D'Eugenio}, {Sun}, {Helton}, {Miles}, {Marley}, {Lew}, {Leisenring}, {Bunker}, {Cargile}, {Carniani}, {Eisenstein}, {Juod{\v{z}}balis}, {Johnson}, {Robertson}, {Tacchella}, {Williams}, \& {Willmer}}]{hainline24b}
{Hainline}, K.~N., {D'Eugenio}, F., {Sun}, F., {et~al.} 2024{\natexlab{b}}, \apj, 975, 31, \dodoi{10.3847/1538-4357/ad76a7}

\bibitem[{{Hainline} {et~al.}(2025){Hainline}, {Helton}, {Miles}, {Leisenring}, {Marley}, {Mukherjee}, {Wogan}, {Bunker}, {Johnson}, {Maiolino}, {Rieke}, {Rinaldi}, {Robertson}, {Sun}, {Tacchella}, {Williams}, \& {Willmer}}]{hainline25}
{Hainline}, K.~N., {Helton}, J.~M., {Miles}, B.~E., {et~al.} 2025, arXiv e-prints, arXiv:2510.00111, \dodoi{10.48550/arXiv.2510.00111}

\bibitem[{{Hainline} {et~al.}(2026){Hainline}, {Eisenstein}, {Whitler}, {Robertson}, {Johnson}, {Jakobsen}, {Puskas}, {Tacchella}, {Helton}, {Wu}, {Arribas}, {Baker}, {Bunker}, {Cameron}, {Carniani}, {Carreira}, {Charlot}, {Chevallard}, {Curtis-Lake}, {D'Eugenio}, {Duan}, {Egami}, {Hausen}, {Ji}, {Looser}, {Maiolino}, {Mengistu}, {Perez-Gonzalez}, {Rieke}, {Rinaldi}, {Sun}, {Trussler}, {Ubler}, {Williams}, {Willmer}, {Willott}, \& {Witstok}}]{hainline2026zgt8}
{Hainline}, K.~N., {Eisenstein}, D.~J., {Whitler}, L., {et~al.} 2026, arXiv e-prints, arXiv:2601.15959, \dodoi{10.48550/arXiv.2601.15959}

\bibitem[{{Jeon} {et~al.}(2026){Jeon}, {Bromm}, {Venditti}, {Finkelstein}, \& {Hsiao}}]{Jeon_PISN2026}
{Jeon}, J., {Bromm}, V., {Venditti}, A., {Finkelstein}, S.~L., \& {Hsiao}, T. Y.-Y. 2026, \apj, 1001, 3, \dodoi{10.3847/1538-4357/ae517d}

\bibitem[{Juri\'{c} {et~al.}(2008)Juri\'{c}, Ivezi\'{c}, Brooks, Lupton, Schlegel, Finkbeiner, Padmanabhan, Bond, Sesar, Rockosi, Knapp, Gunn, Sumi, Schneider, Barentine, Brewington, Brinkmann, Fukugita, Harvanek, Kleinman, Krzesinski, Long, Neilsen, Nitta, Snedden, \& York}]{juric_milky_2008}
Juri\'{c}, M., Ivezi\'{c}, Z., Brooks, A., {et~al.} 2008, The Astrophysical Journal, 673, 864, \dodoi{10.1086/523619}

\bibitem[{{Kirkpatrick} {et~al.}(2021){Kirkpatrick}, {Gelino}, {Faherty}, {Meisner}, {Caselden}, {Schneider}, {Marocco}, {Cayago}, {Smart}, {Eisenhardt}, {Kuchner}, {Wright}, {Cushing}, {Allers}, {Bardalez Gagliuffi}, {Burgasser}, {Gagn{\'e}}, {Logsdon}, {Martin}, {Ingalls}, {Lowrance}, {Abrahams}, {Aganze}, {Gerasimov}, {Gonzales}, {Hsu}, {Kamraj}, {Kiman}, {Rees}, {Theissen}, {Ammar}, {Andersen}, {Beaulieu}, {Colin}, {Elachi}, {Goodman}, {Gramaize}, {Hamlet}, {Hong}, {Jonkeren}, {Khalil}, {Martin}, {Pendrill}, {Pumphrey}, {Rothermich}, {Sainio}, {Stenner}, {Tanner}, {Th{\'e}venot}, {Voloshin}, {Walla}, {W{\k{e}}dracki}, \& {Backyard Worlds: Planet 9 Collaboration}}]{kirkpatrick21}
{Kirkpatrick}, J.~D., {Gelino}, C.~R., {Faherty}, J.~K., {et~al.} 2021, \apjs, 253, 7, \dodoi{10.3847/1538-4365/abd107}

\bibitem[{{Li} {et~al.}(2026){Li}, {Zhang}, {Peng}, {G{\'a}lvez-Ortiz}, {Zhou}, \& {Jones}}]{li26}
{Li}, D.~H., {Zhang}, Z.~H., {Peng}, H.~H., {et~al.} 2026, \mnras, 547, stag227, \dodoi{10.1093/mnras/stag227}

\bibitem[{Luhman {et~al.}(2023)Luhman, Tremblin, Alves~de Oliveira, Birkmann, Baraffe, Chabrier, Manjavacas, Parker, \& Valenti}]{luhman2024}
Luhman, K.~L., Tremblin, P., Alves~de Oliveira, C., {et~al.} 2023, The Astronomical Journal, 167, 5, \dodoi{10.3847/1538-3881/ad0b72}

\bibitem[{{Martis} {et~al.}(2024){Martis}, {Sarrouh}, {Willott}, {Abraham}, {Asada}, {Brada{\v{c}}}, {Brammer}, {Desprez}, {Harshan}, {Muzzin}, {Noirot}, {Rihtar{\v{s}}i{\v{c}}}, {Sawicki}, \& {Strait}}]{martis24}
{Martis}, N.~S., {Sarrouh}, G. T.~E., {Willott}, C.~J., {et~al.} 2024, \apj, 975, 76, \dodoi{10.3847/1538-4357/ad7735}

\bibitem[{{Meisner} {et~al.}(2023){Meisner}, {Leggett}, {Logsdon}, {Schneider}, {Tremblin}, \& {Phillips}}]{atmopp}
{Meisner}, A.~M., {Leggett}, S.~K., {Logsdon}, S.~E., {et~al.} 2023, \aj, 166, 57, \dodoi{10.3847/1538-3881/acdb68}

\bibitem[{{Morrissey} {et~al.}(2026){Morrissey}, {Burgasser}, {de Graaff}, {McConachie}, \& {Brammer}}]{morrissey26}
{Morrissey}, S.~J., {Burgasser}, A.~J., {de Graaff}, A., {McConachie}, I., \& {Brammer}, G. 2026, \aj, 171, 191, \dodoi{10.3847/1538-3881/ae40f1}

\bibitem[{{Naidu} {et~al.}(2022){Naidu}, {Oesch}, {Setton}, {Matthee}, {Conroy}, {Johnson}, {Weaver}, {Bouwens}, {Brammer}, {Dayal}, {Illingworth}, {Barrufet}, {Belli}, {Bezanson}, {Bose}, {Heintz}, {Leja}, {Leonova}, {Marques-Chaves}, {Stefanon}, {Toft}, {van der Wel}, {van Dokkum}, {Weibel}, \& {Whitaker}}]{naidu22}
{Naidu}, R.~P., {Oesch}, P.~A., {Setton}, D.~J., {et~al.} 2022, arXiv e-prints, arXiv:2208.02794, \dodoi{10.48550/arXiv.2208.02794}

\bibitem[{{Naidu} {et~al.}(2026){Naidu}, {Oesch}, {Brammer}, {Weibel}, {Li}, {Matthee}, {Chisolm}, {Pollock}, {Heintz}, {Johnson}, {Shen}, {Hviding}, {Leja}, {Tacchella}, {Ganguly}, {Witten}, {Atek}, {Belli}, {Bose}, {Bouwens}, {Dayal}, {Decarli}, {de Graaff}, {Fudamoto}, {Giovinazzo}, {Greene}, {Illingworth}, {Inoue}, {Kane}, {Labbe}, {Leonova}, {Marques-Chaves}, {Meyer}, {Nelson}, {Roberts-Borsani}, {Schaerer}, {Simcoe}, {Stefanon}, {Sugahara}, {Toft}, {van der Wel}, {van Dokkum}, {Walter}, {Watson}, {Weaver}, \& {Whitaker}}]{naidu26}
{Naidu}, R.~P., {Oesch}, P.~A., {Brammer}, G., {et~al.} 2026, The Open Journal of Astrophysics, 9, 56033, \dodoi{10.33232/001c.156033}

\bibitem[{{Peng} {et~al.}(2011){Peng}, {Ho}, {Impey}, \& {Rix}}]{peng11}
{Peng}, C.~Y., {Ho}, L.~C., {Impey}, C.~D., \& {Rix}, H.-W. 2011, {GALFIT: Detailed Structural Decomposition of Galaxy Images}, Astrophysics Source Code Library.
\newblock \doeprint{1104.010}

\bibitem[{{P{\'e}rez-Gonz{\'a}lez} {et~al.}(2025){P{\'e}rez-Gonz{\'a}lez}, {{\"O}stlin}, {Costantin}, {Melinder}, {Finkelstein}, {Somerville}, {Annunziatella}, {{\'A}lvarez-M{\'a}rquez}, {Colina}, {Dekel}, {Ferguson}, {Li}, {Yung}, {Bagley}, {Boogaard}, {Burgarella}, {Calabr{\`o}}, {Caputi}, {Cheng}, {Dickinson}, {Eckart}, {Giavalisco}, {Gillman}, {Greve}, {Hamed}, {Hathi}, {Hjorth}, {Huertas-Company}, {Kartaltepe}, {Koekemoer}, {Kokorev}, {Labiano}, {Langeroodi}, {Leung}, {Natarajan}, {Papovich}, {Peissker}, {Pentericci}, {Pirzkal}, {Rinaldi}, {van der Werf}, \& {Walter}}]{perezgonzalez25}
{P{\'e}rez-Gonz{\'a}lez}, P.~G., {{\"O}stlin}, G., {Costantin}, L., {et~al.} 2025, \apj, 991, 179, \dodoi{10.3847/1538-4357/adf8c9}

\bibitem[{{Phillips} {et~al.}(2020){Phillips}, {Tremblin}, {Baraffe}, {Chabrier}, {Allard}, {Spiegelman}, {Goyal}, {Drummond}, \& {H{\'e}brard}}]{phillips2020}
{Phillips}, M.~W., {Tremblin}, P., {Baraffe}, I., {et~al.} 2020, \aap, 637, A38, \dodoi{10.1051/0004-6361/201937381}

\bibitem[{{Rihtar{\v{s}}i{\v{c}}} {et~al.}(2026){Rihtar{\v{s}}i{\v{c}}}, {Brada{\v{c}}}, {Desprez}, {Harshan}, {Martis}, {Willott}, {Asada}, {Sarrouh}, {Cornil-Baiotto}, {Biviano}, {Clowe}, {Gonzalez}, {Jones}, {Jude{\v{z}}}, {Kim}, {Lombardi}, {Marchesini}, {Markevitch}, {Markov}, {Noirot}, {Peter}, {Randall}, {Robertson}, {Sawicki}, \& {Tripodi}}]{rihtarsic26}
{Rihtar{\v{s}}i{\v{c}}}, G., {Brada{\v{c}}}, M., {Desprez}, G., {et~al.} 2026, arXiv e-prints, arXiv:2601.22245, \dodoi{10.48550/arXiv.2601.22245}

\bibitem[{{Rothermich} {et~al.}(2024){Rothermich}, {Faherty}, {Bardalez-Gagliuffi}, {Schneider}, {Kirkpatrick}, {Meisner}, {Burgasser}, {Kuchner}, {Allers}, {Gagn{\'e}}, {Caselden}, {Calamari}, {Popinchalk}, {Su{\'a}rez}, {Gerasimov}, {Aganze}, {Softich}, {Hsu}, {Karpoor}, {Theissen}, {Rees}, {Cecilio-Flores-Elie}, {Cushing}, {Marocco}, {Casewell}, {Bickle}, {Hamlet}, {Allen}, {Beaulieu}, {Colin}, {Gantier}, {Gramaize}, {Jalowiczor}, {Kabatnik}, {Kiwy}, {Martin}, {Pendrill}, {Pumphrey}, {Sainio}, {Schumann}, {Stevnbak}, {Sun}, {Tanner}, {Thakur}, {Thevenot}, \& {Wedracki}}]{rothermich2024}
{Rothermich}, A., {Faherty}, J.~K., {Bardalez-Gagliuffi}, D., {et~al.} 2024, \aj, 167, 253, \dodoi{10.3847/1538-3881/ad324e}

\bibitem[{{Ryan} {et~al.}(2011){Ryan}, {Thorman}, {Yan}, {Fan}, {Yan}, {Mechtley}, {Hathi}, {Cohen}, {Windhorst}, {McCarthy}, \& {Wittman}}]{ryan11}
{Ryan}, R.~E., {Thorman}, P.~A., {Yan}, H., {et~al.} 2011, \apj, 739, 83, \dodoi{10.1088/0004-637X/739/2/83}

\bibitem[{{Sanghi} {et~al.}(2023){Sanghi}, {Liu}, {Best}, {Dupuy}, {Siverd}, {Zhang}, {Hurt}, {Magnier}, {Aller}, \& {Deacon}}]{sanghi23}
{Sanghi}, A., {Liu}, M.~C., {Best}, W. M.~J., {et~al.} 2023, \apj, 959, 63, \dodoi{10.3847/1538-4357/acff66}

\bibitem[{{Sarrouh} {et~al.}(2025){Sarrouh}, {Asada}, {Martis}, {Willott}, {Iyer}, {Noirot}, {Muzzin}, {Sawicki}, {Brammer}, {Desprez}, {Rihtar{\v{s}}i{\v{c}}}, {Zabl}, {Abraham}, {Brada{\v{c}}}, {Doyon}, {Antwi-Danso}, {Berek}, {Brown}, {Estrada-Carpenter}, {Favaro}, {Felicioni}, {Forrest}, {Gaspar}, {Gould}, {Gledhill}, {Harshan}, {Jahan}, {Jagga}, {Jude{\v{z}}}, {Marchesini}, {Markov}, {Matharu}, {MacFarland}, {Merchant}, {M{\'e}rida}, {Mowla}, {Myers}, {Omori}, {Pacifici}, {Ravindranath}, {Robbins}, {Strait}, {Sok}, {Tan}, {Tripodi}, {Wilson}, \& {Withers}}]{canucsdr}
{Sarrouh}, G. T.~E., {Asada}, Y., {Martis}, N.~S., {et~al.} 2025, arXiv e-prints, arXiv:2506.21685, \dodoi{10.48550/arXiv.2506.21685}

\bibitem[{{Softich} {et~al.}(2022){Softich}, {Schneider}, {Patience}, {Burgasser}, {Shkolnik}, {Faherty}, {Caselden}, {Meisner}, {Kirkpatrick}, {Kuchner}, {Gagn{\'e}}, {Gagliuffi}, {Cushing}, {Casewell}, {Aganze}, {Hsu}, {Andersen}, {Kiwy}, {Th{\'e}venot}, \& {Backyard Worlds: Planet 9 Collaboration}}]{softich2022}
{Softich}, E., {Schneider}, A.~C., {Patience}, J., {et~al.} 2022, \apjl, 926, L12, \dodoi{10.3847/2041-8213/ac51d8}

\bibitem[{{Tu} {et~al.}(2025){Tu}, {Wang}, {Chen}, \& {Liu}}]{tu25}
{Tu}, Z., {Wang}, S., {Chen}, X., \& {Liu}, J. 2025, \apj, 980, 230, \dodoi{10.3847/1538-4357/adaf9f}

\bibitem[{{Weinberg} {et~al.}(1987){Weinberg}, {Shapiro}, \& {Wasserman}}]{weinberg1987}
{Weinberg}, M.~D., {Shapiro}, S.~L., \& {Wasserman}, I. 1987, \apj, 312, 367, \dodoi{10.1086/164883}

\bibitem[{{Windhorst} {et~al.}(2011){Windhorst}, {Cohen}, {Hathi}, {McCarthy}, {Ryan}, {Yan}, {Baldry}, {Driver}, {Frogel}, {Hill}, {Kelvin}, {Koekemoer}, {Mechtley}, {O'Connell}, {Robotham}, {Rutkowski}, {Seibert}, {Straughn}, {Tuffs}, {Balick}, {Bond}, {Bushouse}, {Calzetti}, {Crockett}, {Disney}, {Dopita}, {Hall}, {Holtzman}, {Kaviraj}, {Kimble}, {MacKenty}, {Mutchler}, {Paresce}, {Saha}, {Silk}, {Trauger}, {Walker}, {Whitmore}, \& {Young}}]{windhorst11}
{Windhorst}, R.~A., {Cohen}, S.~H., {Hathi}, N.~P., {et~al.} 2011, \apjs, 193, 27, \dodoi{10.1088/0067-0049/193/2/27}

\bibitem[{{Wogan} {et~al.}(2025){Wogan}, {Mang}, {Batalha}, {Zahnle}, {Mukherjee}, {Visscher}, {Fortney}, {Marley}, \& {Morley}}]{elfowl25}
{Wogan}, N.~F., {Mang}, J., {Batalha}, N.~E., {et~al.} 2025, Research Notes of the American Astronomical Society, 9, 108, \dodoi{10.3847/2515-5172/add407}

\end{thebibliography}

\end{document}